\documentclass[11pt,a4paper]{article}
\pdfoutput=1 
\usepackage{jheppub} 
\usepackage[T1]{fontenc} % if needed
\usepackage[utf8x]{inputenc}
\usepackage{gensymb}
\usepackage{subfig}
\usepackage{graphicx}
\usepackage{textcomp}
\usepackage{textgreek}
\emergencystretch=\maxdimen
\title{\boldmath Electroluminescence TPCs at the thermal diffusion limit}

\collaboration{The NEXT Collaboration}
\author[a]{C.A.O.~Henriques,}
\author[a]{C.M.B.~Monteiro,}
\author[b]{D.~Gonz\'alez-D\'iaz,}
\author[c]{C.D.R~Azevedo,}
\author[a]{E.D.C.~Freitas,}
\author[a]{R.D.P.~Mano,}
\author[a]{M.R.~Jorge,}
\author[a]{A.F.M.~Fernandes,}
\author[d,e,1]{J.J.~G\'omez-Cadenas,\note[1]{NEXT Co-spokesperson.}}
\author[a,2]{L.M.P.~Fernandes,\note[2]{Corresponding author.}}
\author[f]{C.~Adams,}
\author[g]{V.~\'Alvarez,}
\author[h]{L.~Arazi,}
\author[i]{K.~Bailey,}
\author[j]{F.~Ballester,}
\author[g]{J.M.~Benlloch-Rodr\'{i}guez,}
\author[k]{F.I.G.M.~Borges,}
\author[g]{A.~Botas,}
\author[g]{S.~C\'arcel,}
\author[g]{J.V.~Carri\'on,}
\author[l]{S.~Cebri\'an,}
\author[k]{C.A.N.~Conde,}
\author[g]{J.~D\'iaz,}
\author[m]{M.~Diesburg,}
\author[k]{J.~Escada,}
\author[j]{R.~Esteve,}
\author[g]{R.~Felkai,}
\author[d,e]{P.~Ferrario,}
\author[c]{A.L.~Ferreira,}
\author[d]{J.~Generowicz,}
\author[n]{A.~Goldschmidt,}
\author[f]{R.~Guenette,}
\author[o]{R.M.~Guti\'errez,}
\author[i]{K.~Hafidi,}
\author[p]{J.~Hauptman,}
\author[o]{A.I.~Hernandez,}
\author[b]{J.A.~Hernando~Morata,}
\author[j]{V.~Herrero,}
\author[i]{S.~Johnston,}
\author[q]{B.J.P.~Jones,}
\author[g]{M.~Kekic,}
\author[r]{L.~Labarga,}
\author[g]{A.~Laing,}
\author[m]{P.~Lebrun,}
\author[g]{N.~L\'opez-March,}
\author[o]{M.~Losada,}
\author[f]{J.~Mart\'in-Albo,}
\author[g]{A.~Mart\'inez,}
\author[b,g]{G.~Mart\'inez-Lema,}
\author[q]{A.~McDonald,}
\author[d,q]{F.~Monrabal,}
\author[j]{F.J.~Mora,}
\author[g]{J.~Mu\~noz Vidal,}
\author[g]{M.~Musti,}
\author[g]{M.~Nebot-Guinot,}
\author[g]{P.~Novella,}
\author[q,3]{D.R.~Nygren,\note[3]{NEXT Co-spokesperson.}}
\author[g]{B.~Palmeiro,}
\author[m]{A.~Para,}
\author[g,4]{J.~P\'erez,\note[4]{Now at Laboratorio Subterr\'aneo de Canfranc, Spain.}}
\author[q]{F.~Psihas,}
\author[g]{M.~Querol,}
\author[g]{J.~Renner,}
\author[i]{J.~Repond,}
\author[i]{S.~Riordan,}
\author[s]{L.~Ripoll,}
\author[g]{J.~Rodr\'iguez,}
\author[q]{L.~Rogers,}
\author[g]{C.~Romo-Luque,}
\author[k]{F.P.~Santos,}
\author[a]{J.M.F. dos~Santos,}
\author[g,h]{A.~Sim\'on,}
\author[t,5]{C.~Sofka,\note[5]{Now at University of Texas at Austin, USA.}}
\author[g]{M.~Sorel,}
\author[t]{T.~Stiegler,}
\author[j]{J.F.~Toledo,}
\author[d]{J.~Torrent,}
\author[c]{J.F.C.A.~Veloso,}
\author[t]{R.~Webb,}
\author[t,6]{J.T.~White,\note[6]{Deceased.}}
\author[g]{N.~Yahlali}

\emailAdd{pancho@gian.fis.uc.pt}
\affiliation[a]{
LIBPhys, Physics Department, University of Coimbra, Rua Larga, Coimbra, 3004-516, Portugal}
\affiliation[b]{
Instituto Gallego de F\'isica de Altas Energ\'ias, Univ.\ de Santiago de Compostela, Campus sur, R\'ua Xos\'e Mar\'ia Su\'arez N\'u\~nez, s/n, Santiago de Compostela, E-15782, Spain}
\affiliation[c]{
Institute of Nanostructures, Nanomodelling and Nanofabrication (i3N), Universidade de Aveiro, Campus de Santiago, Aveiro, 3810-193, Portugal}
\affiliation[d]{
Donostia International Physics Center (DIPC), Paseo Manuel Lardizabal, 4, Donostia-San Sebastian, E-20018, Spain}
\affiliation[e]{
Ikerbasque, Basque Foundation for Science, Bilbao, E-48013, Spain}
\affiliation[f]{
Department of Physics, Harvard University, Cambridge, MA 02138, USA}
\affiliation[g]{
Instituto de F\'isica Corpuscular (IFIC), CSIC \& Universitat de Val\`encia, Calle Catedr\'atico Jos\'e Beltr\'an, 2, Paterna, E-46980, Spain}
\affiliation[h]{
Nuclear Engineering Unit, Faculty of Engineering Sciences, Ben-Gurion University of the Negev, P.O.B. 653, Beer-Sheva, 8410501, Israel}
\affiliation[i]{
Argonne National Laboratory, Argonne, IL 60439, USA}
\affiliation[j]{
Instituto de Instrumentaci\'on para Imagen Molecular (I3M), Centro Mixto CSIC - Universitat Polit\`ecnica de Val\`encia, Camino de Vera s/n, Valencia, E-46022, Spain}
\affiliation[k]{
LIP, Department of Physics, University of Coimbra, Coimbra, 3004-516, Portugal}
\affiliation[l]{
Laboratorio de F\'isica Nuclear y Astropart\'iculas, Universidad de Zaragoza, Calle Pedro Cerbuna, 12, Zaragoza, E-50009, Spain}
\affiliation[m]{
Fermi National Accelerator Laboratory, Batavia, IL 60510, USA}
\affiliation[n]{
Lawrence Berkeley National Laboratory (LBNL), 1 Cyclotron Road, Berkeley, CA 94720, USA}
\affiliation[o]{
Centro de Investigaci\'on en Ciencias B\'asicas y Aplicadas, Universidad Antonio Nari\~no, Sede Circunvalar, Carretera 3 Este No.\ 47 A-15, Bogot\'a, Colombia}
\affiliation[p]{
Department of Physics and Astronomy, Iowa State University, 12 Physics Hall, Ames, IA 50011-3160, USA}
\affiliation[q]{
Department of Physics, University of Texas at Arlington, Arlington, TX 76019, USA}
\affiliation[r]{
Departamento de F\'isica Te\'orica, Universidad Aut\'onoma de Madrid, Campus de Cantoblanco, Madrid, E-28049, Spain}
\affiliation[s]{
Escola Polit\`ecnica Superior, Universitat de Girona, Av.~Montilivi, s/n, Girona, E-17071, Spain}
\affiliation[t]{
Department of Physics and Astronomy, Texas A\&M University, College Station, TX 77843-4242, USA}

\abstract{The NEXT experiment aims at searching for the hypothetical neutrinoless double-beta decay from the ${}^{136}$Xe isotope using a high-purity xenon TPC. Efficient discrimination of the events through pattern recognition of the topology of primary ionisation tracks is a major requirement for the experiment. However, it is limited by the diffusion of electrons. It is known that the addition of a small fraction of a molecular gas to xenon reduces electron diffusion. On the other hand, the electroluminescence (EL) yield drops and the achievable energy resolution may be compromised. We have studied the effect of adding several molecular gases to xenon (CO${}_{2}$, CH${}_{4}$ and CF${}_{4}$) on the EL yield and energy resolution obtained in a small prototype of driftless gas proportional scintillation counter. We have compared our results on the scintillation characteristics (EL yield and energy resolution) with a microscopic simulation, obtaining the diffusion coefficients in those conditions as well. Accordingly, electron diffusion may be reduced from about 10 mm/$\sqrt{\mathrm{m}}$ for pure xenon down to 2.5 mm/$\sqrt{\mathrm{m}}$ using additive concentrations of about 0.05\%, 0.2\% and 0.02\% for CO${}_{2}$, CH${}_{4}$ and CF${}_{4}$, respectively. Our results show that CF${}_{4}$ admixtures present the highest EL yield in those conditions, but very poor energy resolution as a result of huge fluctuations observed in the EL formation. CH${}_{4}$ presents the best energy resolution despite the EL yield being the lowest. The results obtained with xenon admixtures are extrapolated to the operational conditions of the NEXT-100 TPC. CO${}_{2}$ and CH${}_{4}$ show potential as molecular additives in a large xenon TPC. While CO${}_{2}$ has some operational constraints, making it difficult to be used in a large TPC, CH${}_{4}$ shows the best performance and stability as molecular additive to be used in the NEXT-100 TPC, with an extrapolated energy resolution of 0.4\% at 2.45 MeV for concentrations below 0.4\%, which is only slightly worse than the one obtained for pure xenon. We demonstrate the possibility to have an electroluminescence TPC operating very close to the thermal diffusion limit without jeopardizing the TPC performance, if CO${}_{2}$ or CH${}_{4}$ are chosen as additives.}

\begin{document} 
\maketitle
\flushbottom

\section{Introduction}
\label{sec:intro}
High-pressure xenon (HPXe) time projection chambers (TPCs) are increasingly used in applications for rare-event detection such as double-beta decay (DBD) and double-electron capture (DEC), with or without neutrino emission, as well as directional dark matter (DDM)~\cite{r1,r2,r3,r4,r5,r6,r7}. One such experiment is NEXT, to search for neutrinoless double-beta ($\beta\beta0\nu$) decay~\cite{r1}.

Gaseous xenon TPCs offer important advantages when compared to liquid xenon and double phase xenon TPCs~\cite{r8,r9,r10,r11,r12,r13,r14}. The impact of background depends strongly on the achieved energy resolution, which is much better in the gas phase. Furthermore, interaction of rare events in the gas will allow a determination of the event topological signature, as demonstrated for DBD and DEC detection~\cite{r15, r16, r5}, in contrast with interaction in the liquid, where the extremely reduced dimensions of the primary ionisation trail rules out any possible topology-based pattern recognition.

In particular, optical TPCs based on electroluminescence (EL) amplification of the primary ionisation signal are the most competitive alternatives to charge avalanche amplification TPCs, making it possible to reach energy resolutions as low as 0.7\% FWHM at 2.5 MeV (Q${}_{\beta\beta}$ for $\beta\beta0\nu$ decay of ${}^{136}$Xe), as demonstrated for a 1 kg-scale prototype~\cite{r17}, a factor 2 to 4 better than the energy resolution achieved in TPCs based on charge avalanche readout~\cite{r18}. In addition, EL readout through a photosensor has the advantage of electrically decoupling the amplification region from the photosensor, rendering more immunity to electronic noise, radio-frequency pickup and high voltage discharges.

The EL yield depends on the detector geometry. Absolute values have been measured in uniform electric field detectors~\cite{r19, r20, r21} and in  modern micropatterned electron multipliers, like GEM, THGEM, MHSP and Micromegas~\cite{r22,r23,r24}. While statistical fluctuations in the scintillation produced in charge avalanches are dominated by the variance of the total number of electrons produced, the statistical fluctuations in the EL produced in uniform electric fields below the onset of electron multiplication are negligible when compared to those associated with the primary ionisation formation~\cite{r25}. An efficient way of background discrimination in $\beta\beta0\nu$ decay experiments is based on the energy deposited in the gas. In this case, excellent energy resolution is needed for efficient background rejection, hence TPCs based on EL produced in uniform electric fields represents the best option.

The topological signature of the ionisation trail in the gas is compromised by the large electron diffusion in xenon resulting from inefficient energy loss in elastic collisions with xenon atoms, in particular for low electric fields (few tens of V/cm/bar) used to drift the primary ionisation cloud towards the signal amplification region, and for large drift distances~\cite{r26}. By adding a molecular gas, like CO${}_{2}$, CH${}_{4}$ or CF${}_{4}$, to pure xenon, new molecular degrees of freedom from vibrational and rotational states are made available for electron energy transfer in inelastic collisions, and the energy distribution of the ionisation electron cloud in the drift region tends to build up around the energy of the first vibrational level, typically $\sim$ 0.1 eV, even in the presence of sub-percent concentrations of the additive. At these mildly supra-thermal energies, electron diffusion is considerably reduced. 

The presence of molecular species in a noble gas was believed to dramatically reduce the EL yield~\cite{r27}. If an electron has a significant probability of colliding with a molecular impurity before it obtains from the electric field sufficient energy to excite a noble gas atom, it may lose part of its energy without producing EL photons. Besides electron cooling, additional losses of scintillation originate from excimer quenching, photo-absorption and electron attachment~\cite{r32}. The reduction of the EL yield depends on the amount and type of impurity present in the gas. Recent experimental studies with Xe-CO${}_{2}$ and Xe-CH${}_{4}$ mixtures, for different concentrations of the additives, have shown that the EL yield reduction is not as drastic as previously believed~\cite{r28}.

In the HPXe TPC used by the NEXT collaboration, electron diffusion may be as high as 10 mm/$\sqrt{\mathrm{m}}$, making the pattern recognition of the primary ionisation trail very difficult at the 1 m drift scale. Hence, a campaign designed to systematically add several molecular gases to xenon, at minute concentration levels, has started within the scope of the NEXT experiment. The aim is to find a suitable mixture able to reduce diffusion and improve the topological discrimination of the events, without compromising significantly the performance of the detector in terms of EL yield and energy resolution. 

Detailed studies with Xe-CO${}_{2}$ mixtures, performed at atmospheric pressure, have shown a tolerable 35\% reduction in the EL yield for 0.04\% concentration of CO${}_{2}$ relative to the EL yield obtained in pure Xe~\cite{r29}. On the other hand, simulation results show that the same amount of CO${}_{2}$ would reduce the diffusion coefficient from 10 to about 3 mm/$\sqrt{\mathrm{m}}$~\cite{r28}. The energy resolution obtained with CO${}_{2}$ concentrations up to 0.04\% is not significantly affected, given that the contribution of the statistical fluctuations associated to EL production is significantly lower than the Fano factor contribution. The intrinsic energy resolution deteriorates with increasing CO${}_{2}$ concentration~\cite{r29}. One problem found during long term operation of CO${}_{2}$ mixtures is the instability coming from CO${}_{2}$ adsorption in getters and subsequent formation of CO. This may be a critical problem for a large chamber operating over long periods of time. 

In this paper two other molecular gases, CH${}_{4}$ and CF${}_{4}$, are investigated using the same driftless GPSC prototype. The paper compares the performance of the detector for the different additives investigated, and different additive concentrations, at atmospheric pressure, in particular its effect on the EL yield and energy resolution.

\section{Experimental setup}
\label{sec:setup}

\begin{figure}[tbp]
\centering 
\includegraphics[width=\textwidth,height=\textheight,keepaspectratio]{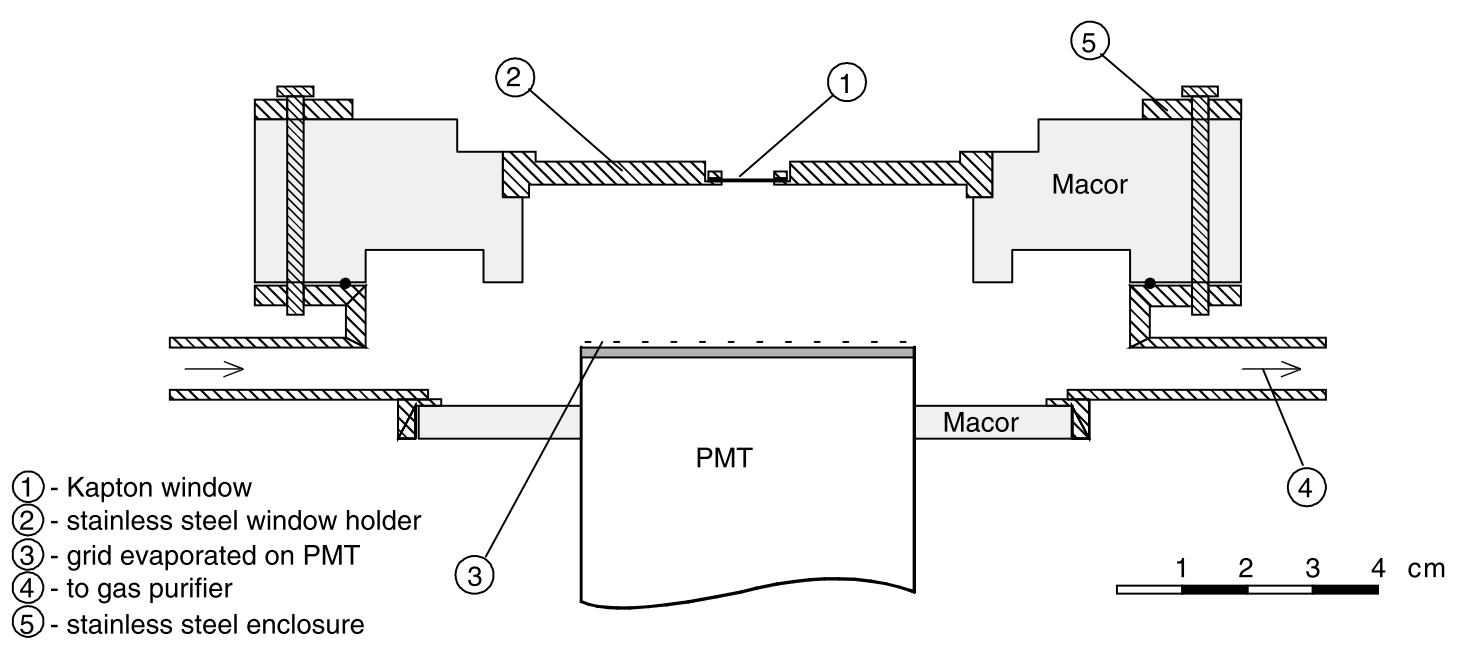}
\caption{\label{fig:detector} Scheme of the driftless GPSC used in this work~\cite{r30}. A PMT is used as VUV photosensor and the gas is continuously purified through SAES St707 getters.}
\end{figure}

The studies described in this paper were performed in a gas proportional scintillation counter (GPSC) without drift region, depicted in figure~\ref{fig:detector}, already used in previous studies~\cite{r29,r30,r31}. The driftless design of the detector allows a study of the influence of molecular additives on the secondary scintillation produced upon x-ray interactions, minimizing the effects that may arise in a typical GPSC as a result of the electron drift through the absorption/conversion region. The scintillation region is 2.5 cm long and is delimited by a Kapton window (8 mm diameter), aluminised on the inner side, and by the quartz window of the photomultiplier tube (PMT), vacuum-evaporated with a chromium grid (strips of 100 {\textmu}m width and 1000 {\textmu}m spacing) and electrically connected to the PMT photocathode. The EL electric field is established by applying a negative high voltage to the detector window and its stainless steel holder, which are insulated from the detector body by a ceramic material (Macor), while the detector body, the chromium grid at the PMT window and the photocathode are kept at 0 V. The reduced electric field inside the detector is set below the gas ionisation threshold so that EL photons can be produced in the scintillation region without charge multiplication.

The performance of the detector was assessed using an x-ray beam from a ${}^{55}$Fe radioactive source, collimated to a 2 mm diameter, irradiating the GPSC window along the detector axis. Only 5.9 keV x-rays (Mn ${K}_{\alpha}$ line) interact in the gas since 6.4 keV x-rays (Mn ${K}_{\beta}$ line) are absorbed by a chromium film. The ionisation electrons released by the interaction of 5.9 keV x-rays are accelerated as they cross the scintillation region, exciting the noble gas atoms and inducing the emission of EL photons. The amount of EL photons is more than 3 orders of magnitude larger than the amount of primary scintillation photons also produced by the x-ray interaction in the gas medium~\cite{r33}. A PMT is used as photosensor for the vacuum ultraviolet (VUV) photons produced in the gas. Pulses produced in the PMT are subsequently shaped, amplified and analysed with a multi-channel analyser (MCA). 

The GPSC is coupled to a gas re-circulation system in order to continuously purify the xenon gas or xenon-additive mixture through SAES St707 getters. A residual gas analyser (RGA) is connected to the gas system through a heated capillary in order to reduce pressure down to about ${10}^{-5}$ mbar, required for the RGA safe operation. The RGA volume is coupled to a vacuum pumping system to extract the gas that continuously enters through the capillary. The RGA is particularly important for the studies with molecular species added to the xenon gas since it provides a real-time direct measurement of the additive concentrations. Two different volumes were added to the system, one filled with pure xenon and the other one with the molecular additive, in order to calibrate the RGA, as described in ref.~\cite{r29}. The pressure in each volume is read by accurate capacitive pressure gauges. In order to avoid any dependence on pressure, the RGA calibration and the detector operation were carried out at the same total pressure for pure xenon and xenon admixtures. Total pressures of 1.13, 1.25 and 1.24 bar were used for CO${}_{2}$, CH${}_{4}$ and CF${}_{4}$ additives, respectively. The EL measurements were performed after stabilisation of the partial pressure measurement in the RGA. 

Before setting each mixture, a measurement of the background is performed with the RGA in the GPSC filled with pure xenon, with the getters operating at 250\textdegree C, in order to ensure maximum xenon purity. The origin of the background most likely comes from degassing of the RGA chamber or capillary, or from gases entering the chamber through the pumps. This background is afterwards subtracted from the RGA reading, once the mixture is prepared. For each molecular additive, a calibration was performed by plotting the additive concentration measured on the RGA as a function of the initial concentration estimated from the pressure measurements on the capacitive gauges connected to the two volumes. Within the concentration range used for each additive, the calibration process showed a good linear correlation between the concentration measured on the RGA and the estimated one, from which the desired calibration factor was obtained.

The gas was purified by hot getters operated at a specific temperature. For CH${}_{4}$ and CF${}_{4}$, getters were operated at 120\textdegree C, which is enough to maintain the gas purity. At this temperature, CO${}_{2}$ is absorbed by getters and subsequently CO is produced, escaping into the gas. For this reason, in this case the temperature of the getters was set to 80\textdegree C in order to minimize the amount of CO produced. The EL yield obtained for pure xenon drops slightly when getters are cooled from 250\textdegree C down to 80\textdegree C, but only after several days of operation.

\section{Method}
\label{sec:method}

The response of a driftless GPSC depends on the x-ray penetration depth since the number of EL photons produced by the x-ray interaction is related to the distance travelled by the primary electron cloud across the scintillation region. Consequently, the total scintillation spectrum, obtained when integrating over the transit time of the electron cloud (as recorded with our preamplifier/MCA chain), is the convolution of a Gaussian with an exponential function towards the low-energy region, since the x-ray interaction probability follows an exponential law. Provided the absorption length for 5.9 keV x-rays in 1 bar of xenon, about 2.5 mm, is still significant when compared to the 25 mm thick EL region, the full absorption peak in the pulse-height distribution deviates from a Gaussian shape, presenting a tail towards the left side corresponding to lower amplitudes.

The shape of the scintillation spectrum may change when molecular gases are added to pure xenon due to several processes, like quenching, photo-absorption and dissociative attachment. The probability of these effects depends on the type and concentration of the additive. In our previous work on Xe-CO${}_{2}$ mixtures~\cite{r29}, the intrinsic response of the GPSC was estimated by decomposition of the full absorption peak into a sum of several Gaussian functions, corresponding to x-ray interactions at successive depths. They have decreasing areas according to the exponential absorption law for 5.9 keV x-rays, with the same relative FWHM. Their centroid is given by the integration of the EL produced uniformly along the electron path and weighted by the solid angle subtended by the PMT photocathode relative to each point of that path~\cite{r29}. The pulse amplitude and energy resolution were taken from the centroid and FWHM of the rightmost Gaussian curve, corresponding to x-ray interactions near the window (zero penetration), or equivalently to an electron path length of 2.5 cm. However, later studies performed with Xe-CF${}_{4}$ mixtures showed for the higher CF${}_{4}$ concentrations a right-tailed pulse-height distribution instead of a left-tailed one. This effect is mostly attributed to the high electron attachment taking place in such mixtures; primary electron clouds resulting from x-rays absorbed farther away from the photosensor are more affected by attachment during their longer drift, resulting in a smaller amount of collected photons when compared to those produced by x-rays interacting more deeply in the detector, in contrast to what was observed in Xe-CO${}_{2}$ mixtures.

\begin{figure}[tbp]
\centering
\includegraphics[width=7.4cm,height=\textheight,keepaspectratio]{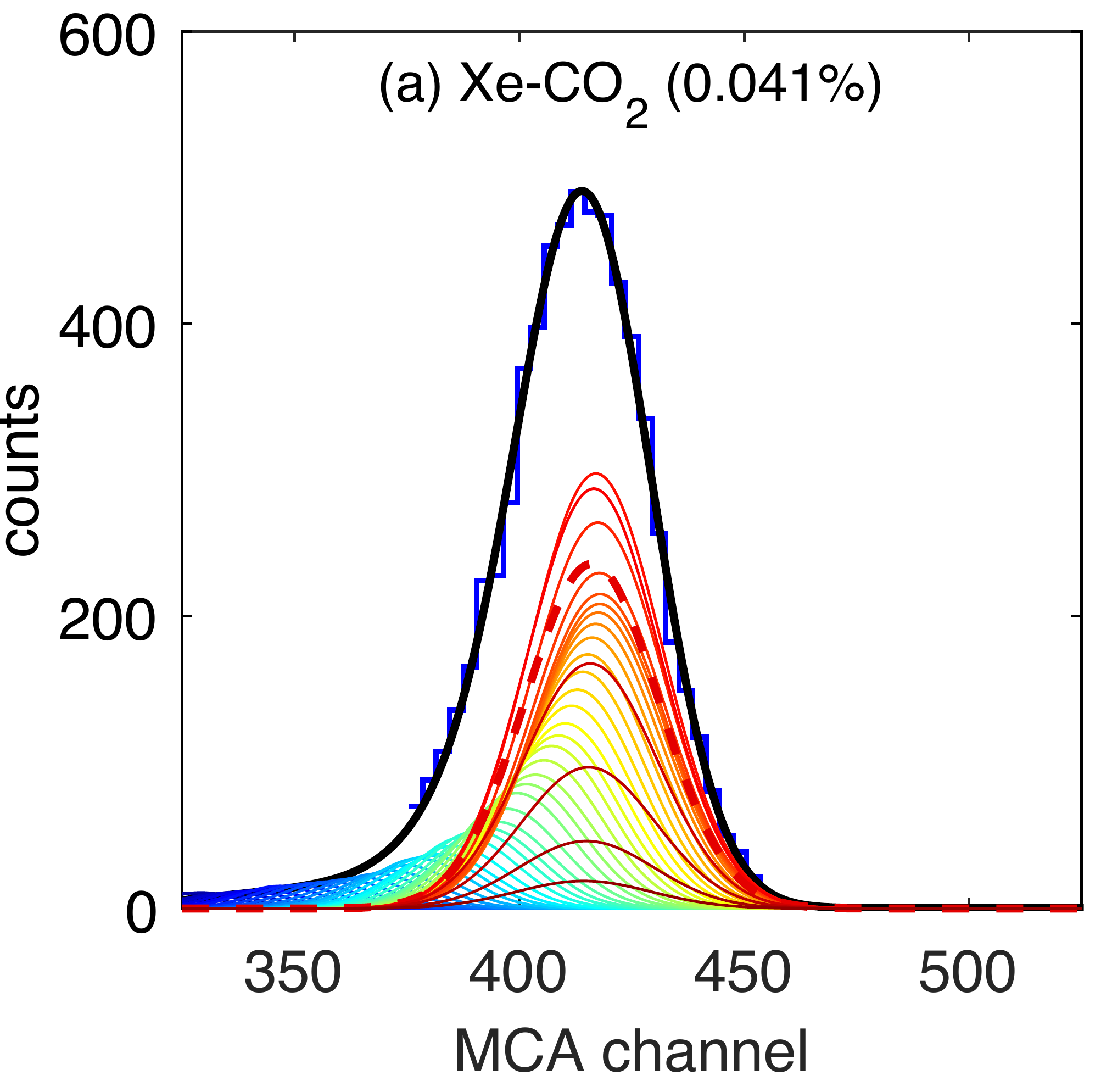}
\includegraphics[width=7.4cm,height=\textheight,keepaspectratio]{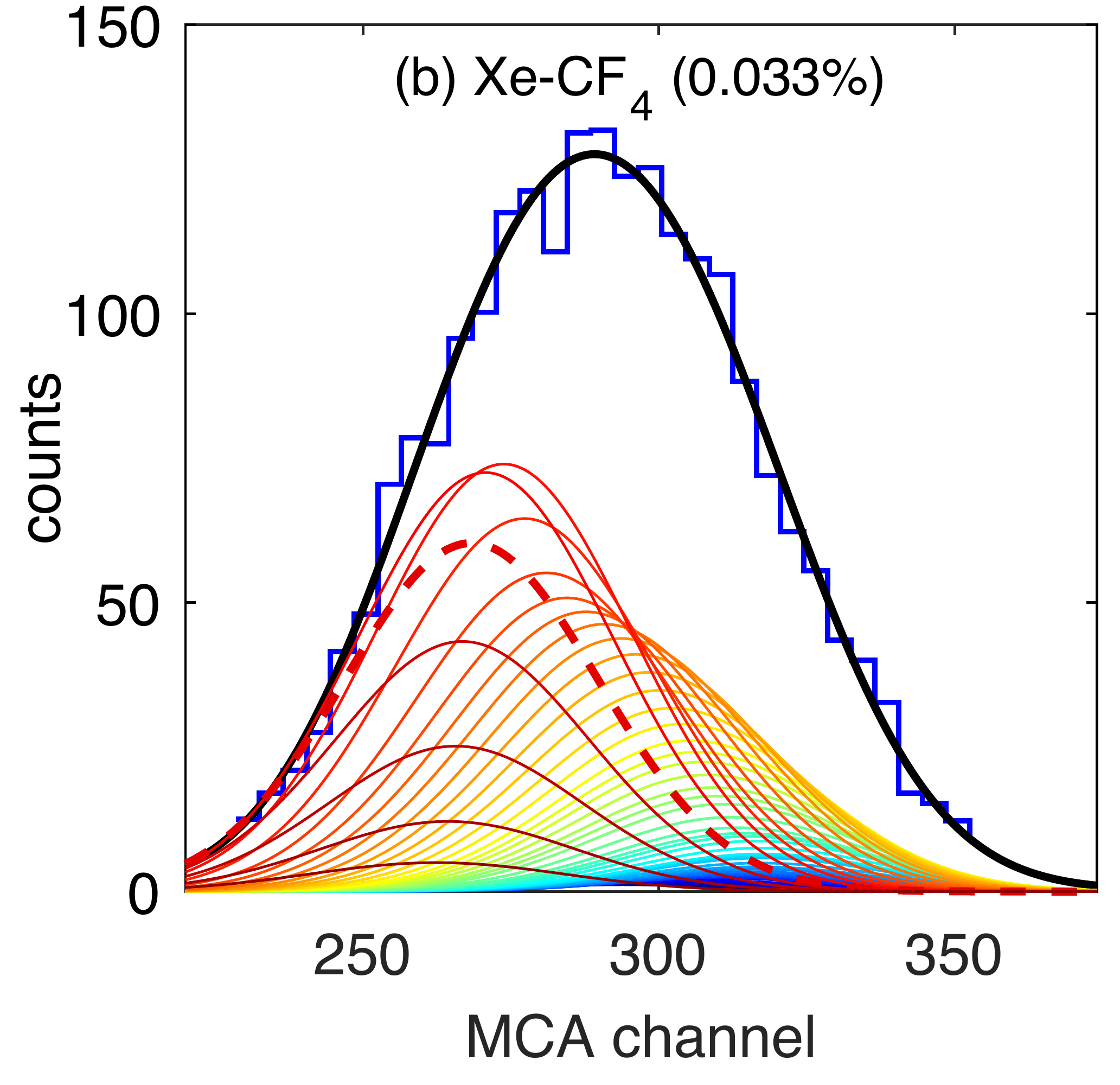}
\caption{\label{fig:method}Scintillation spectra obtained with an MCA for 5.9 keV x-rays absorbed in the GPSC for: (a) a mixture of Xe with 0.041\% of CO${}_{2}$, at 1.15 bar and $E/p=$ 3.1 kV/cm/bar; (b) a mixture of Xe with 0.033\% of CF${}_{4}$, at 1.20 bar and $E/p=$ 2.8 kV/cm/bar. The response function (in black) is a fit to the to the experimental distribution (in dark blue), resulting from the sum of 5000 Gaussian curves (solid thin lines from red to blue, not at vertical scale). The dashed curve is the Gaussian corresponding to 2.5 cm path length, from which the corrected energy resolution is obtained. Corrected values of 8.3\% for Xe-CO${}_{2}$ and 18.7\% for Xe-CF${}_{4}$ were obtained, to be compared to uncorrected values (black curve) of 8.7\% and 21.6\%, respectively. The estimated attachment probability is 7\% for the Xe-CO${}_{2}$ mixture and 80\% for the Xe-CF${}_{4}$ mixture.}
\end{figure}

In figure~\ref{fig:method} we present typical scintillation spectra for 5.9 keV x-rays absorbed in the driftless GPSC in two different cases: one mixture with low attachment probability, Xe with 0.041\% of CO${}_{2}$ (a), and one mixture with high attachment probability, Xe with 0.033\% of CF${}_{4}$ (b). The Xe-CO${}_{2}$ distribution is left-tailed as a result of the x-ray penetration depth in the driftless detector. The Xe-CF${}_{4}$ distribution is right-tailed as a result of the high attachment probability associated to the x-ray penetration effect, presenting much worse energy resolution. Therefore, a more complete decomposition method has to be used to include the effect of electron attachment.

To disentangle the effect of different x-ray penetration depths in the detector, the scintillation spectrum is fitted to the sum of a large number of Gaussian curves (5000) corresponding to different x-ray penetration depths. These Gaussian curves have the same relative FWHM but the areas and centroids follow trends (as a function of the x-ray interaction depth) different from those assumed in~\cite{r29}. To determine those trends, special runs took place in which we used a Lecroy WaveRunner 610Zi digital oscilloscope to observe directly the PMT waveforms using 50-${\Omega}$ DC coupling, in the absence of pre-amplification. Pulses are organized according to their duration, in intervals of 40 ns, which is linearly related to the electron cloud transit time (and, thus, to the x-ray penetration depth) owing to the uniform electric field in our setup. In this way, we obtain the scintillation spectrum for each x-ray interaction depth, which has a Gaussian shape, as expected. The area and centroid position of each Gaussian obtained in this way, as a function of the x-ray interaction depth, are later used in the fitting procedure sketched in figure~\ref{fig:method}. Applying this method to pure Xe and to a Xe-CF${}_{4}$ mixture with known attachment probability, we are able to infer the effect of attachment from the relation between the respective centroids obtained at each interaction depth, being able then to extrapolate the centroid distribution for any other attachment probability. The detailed method, that is essential for correctly interpreting the data at the highest CF${}_{4}$ concentrations, is beyond the scope of this publication.

Finally, the function resulting from the sum of the 5000 Gaussians is fitted to the scintillation spectrum leaving as free parameters the relative FWHM and the centroid of the Gaussian corresponding to zero penetration. The EL yield and energy resolution are taken from the centroid and the relative FWHM of the zero-penetration Gaussian.

\section{Experimental results}
\label{sec:results}

The performance of the driftless GPSC was investigated for xenon with different admixtures of CO${}_{2}$, CH${}_{4}$ and CF${}_{4}$. The behaviour of the EL yield and the energy resolution as a function of the reduced electric field across the GPSC scintillation region was investigated for all mixtures considered. The PMT (EMI D676QB) was operated at a fixed bias voltage of 850 V. The EL yield obtained was compared to simulation results for each mixture, following the approach introduced in~\cite{r32}, and the electron diffusion then obtained within the same framework, that is based on Magboltz~\cite{r39}. The intrinsic energy resolution was estimated following the method described in the previous section. A compromise between low diffusion and good EL performance is made in order to select the best concentration for each additive. Finally, the advantages of each molecular additive are discussed in order to give a hint of the best candidate to be used in the NEXT experiment. 

\subsection{EL yield}

Absolute values of the reduced EL yield ($Y/p$, where $p$ is the gas pressure) were determined by normalization of the pulse amplitude obtained for pure xenon to the ones in~\cite{r19}. The same normalization constant was then used to obtain the remaining EL yield curves for the different mixtures investigated, given that the scintillation spectrum in the PMT band is expected to remain unchanged (PMT and electronic gains are fixed). 

\begin{figure}[tbp]
\centering
\includegraphics[width=7.4cm,height=\textheight,keepaspectratio]{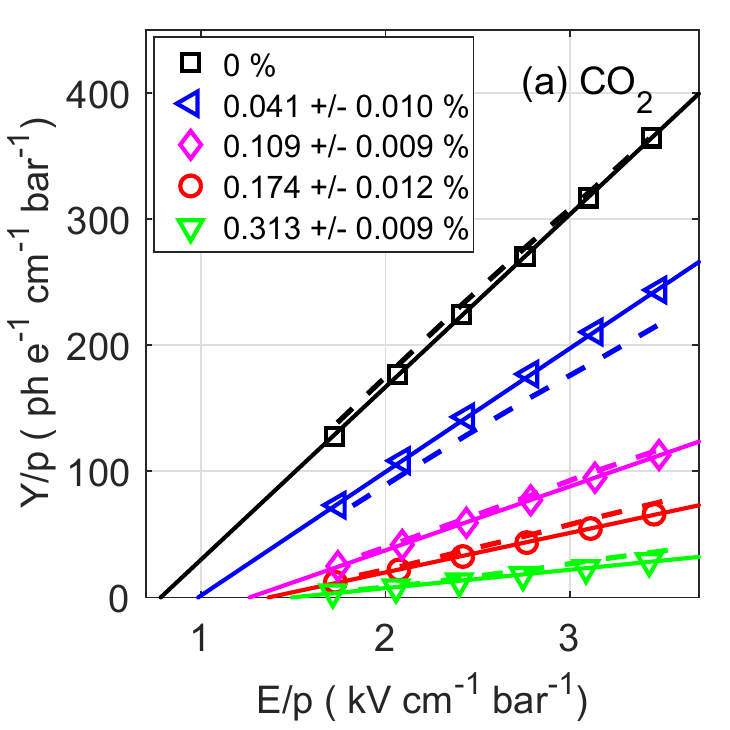} 
\includegraphics[width=7.4cm,height=\textheight,keepaspectratio]{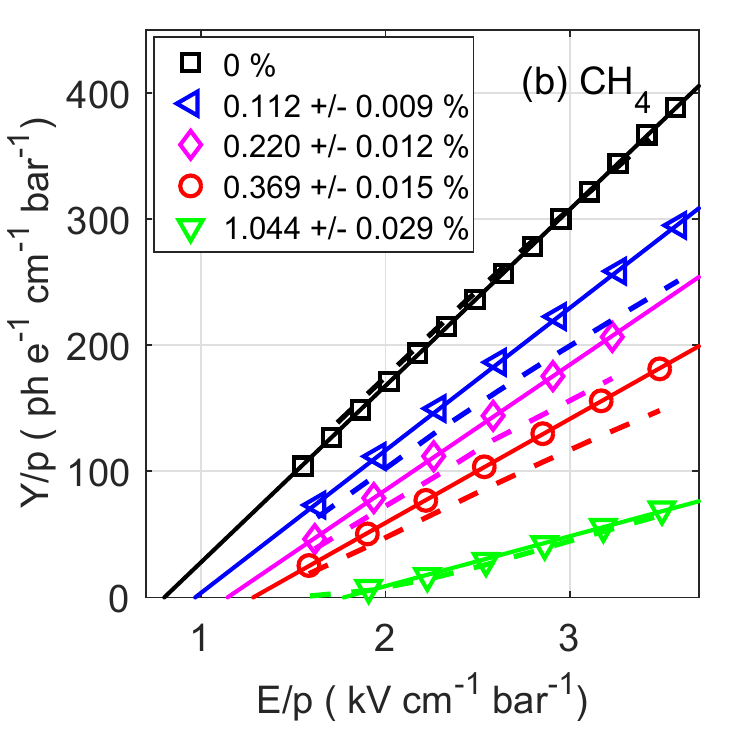} 
\includegraphics[width=7.4cm,height=\textheight,keepaspectratio]{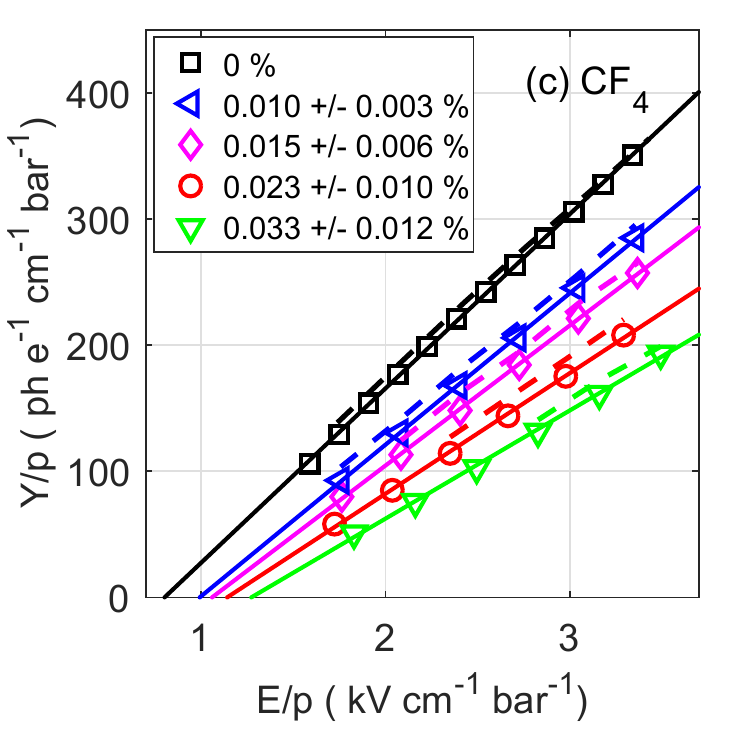} 
\caption{\label{fig:yield} Reduced EL yield obtained for 5.9 keV x-rays, $Y/p$, as a function of the reduced electric field, $E/p$, for different concentrations of molecular gases added to pure xenon: (a) CO${}_{2}$; (b) CH${}_{4}$; (c) CF${}_{4}$. Total pressures of 1.13, 1.25 and 1.24 bar were used, in average, for Xe-CO${}_{2}$, Xe-CH${}_{4}$ and Xe-CF${}_{4}$ mixtures, respectively. Solid lines show linear fits to the data, while dashed lines are simulation values obtained with the code developed in~\cite{r32}.}
\end{figure}

Figure~\ref{fig:yield} presents $Y/p$ as a function of the reduced electric field ($E/p$) applied to the scintillation region, for different concentrations of several molecular gases added to pure xenon: CO${}_{2}$ (a), CH${}_{4}$ (b), and CF${}_{4}$ (c). The solid lines are fits to the data points, while the dashed lines are results from the simulation package introduced in~\cite{r32}. As seen, the reduced EL yield exhibits the typical linear dependence on $E/p$ observed in pure noble gases also in the presence of molecular additives. For the three different additives, the EL yield decreases as the additive concentration increases, for the same $E/p$ value. For a typical reduced electric field of 2.5 kV/cm/bar in the TPC, an EL drop of 50\% relative to the one obtained in pure xenon (black lines) is obtained for different concentrations of the three additives: about 0.05\% for CO${}_{2}$, 0.3\% for CH${}_{4}$ and 0.02\% for CF${}_{4}$. 

Figure~\ref{fig:yield} shows that the EL threshold (minimum $E/p$ value that allows EL multiplication) rises by increasing the amount of additive, while the slope of $Y/p$ against $E/p$ decreases. There are several effects responsible for the variation of the EL yield with the amount of additive. Upon colliding with a molecule, the electron loses energy to rotational and vibrational excited states, reducing its average energy. Electron cooling seems to be very efficient, as seen from the increase of the EL threshold in figure~\ref{fig:yield}, while the scintillation drop remains acceptable up to additive concentrations around 0.1\% for CO${}_{2}$, 0.4\% for CH${}_{4}$, and 0.02\% for CF${}_{4}$. A compromise between electron cooling and excimer scintillation must clearly exist. Although the increase of the EL threshold could in principle be compensated by increasing the electric field, other effects exist that change also the slope of the trends, and cannot be compensated. One of these effects is the excimer quenching, which reduces the probability of scintillation. According to simulation, this effect is dominant for CH${}_{4}$ and negligible for CF${}_{4}$, explaining the smaller variations in the $Y/p$ slope observed for CF${}_{4}$. Another effect is the gas transparency to VUV light, which would be very limiting for CO${}_{2}$ in larger prototypes, despite not contributing at this detector scale. Additional EL losses result from dissociative attachment, which increases with increasing additive concentration. This is the dominant effect for CF${}_{4}$. 

\subsection{Energy resolution}

\begin{figure}[tbp]
\centering 
\includegraphics[width=7.4cm,height=\textheight,keepaspectratio]{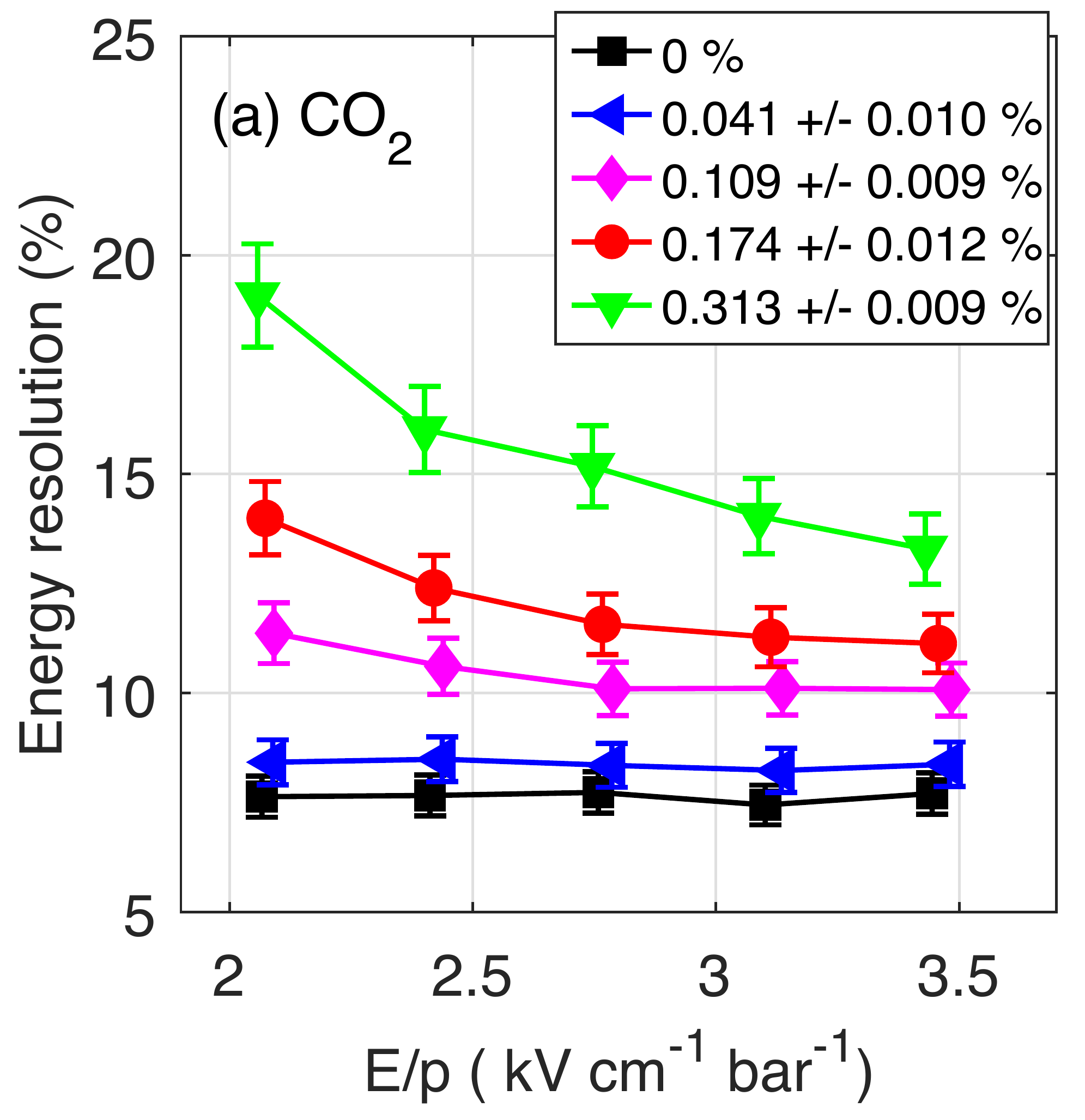}
\includegraphics[width=7.4cm,height=\textheight,keepaspectratio]{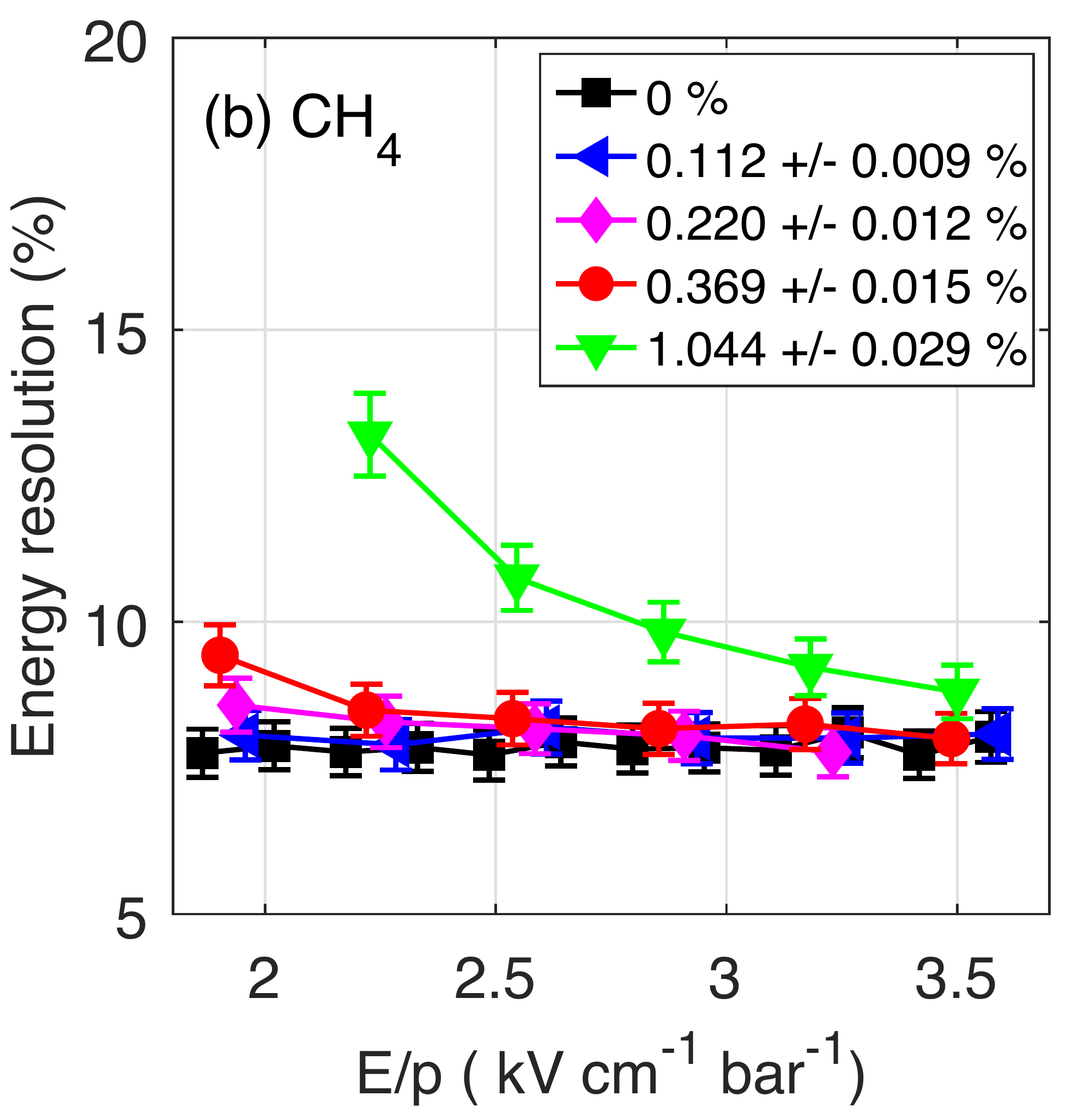}
\includegraphics[width=7.4cm,height=\textheight,keepaspectratio]{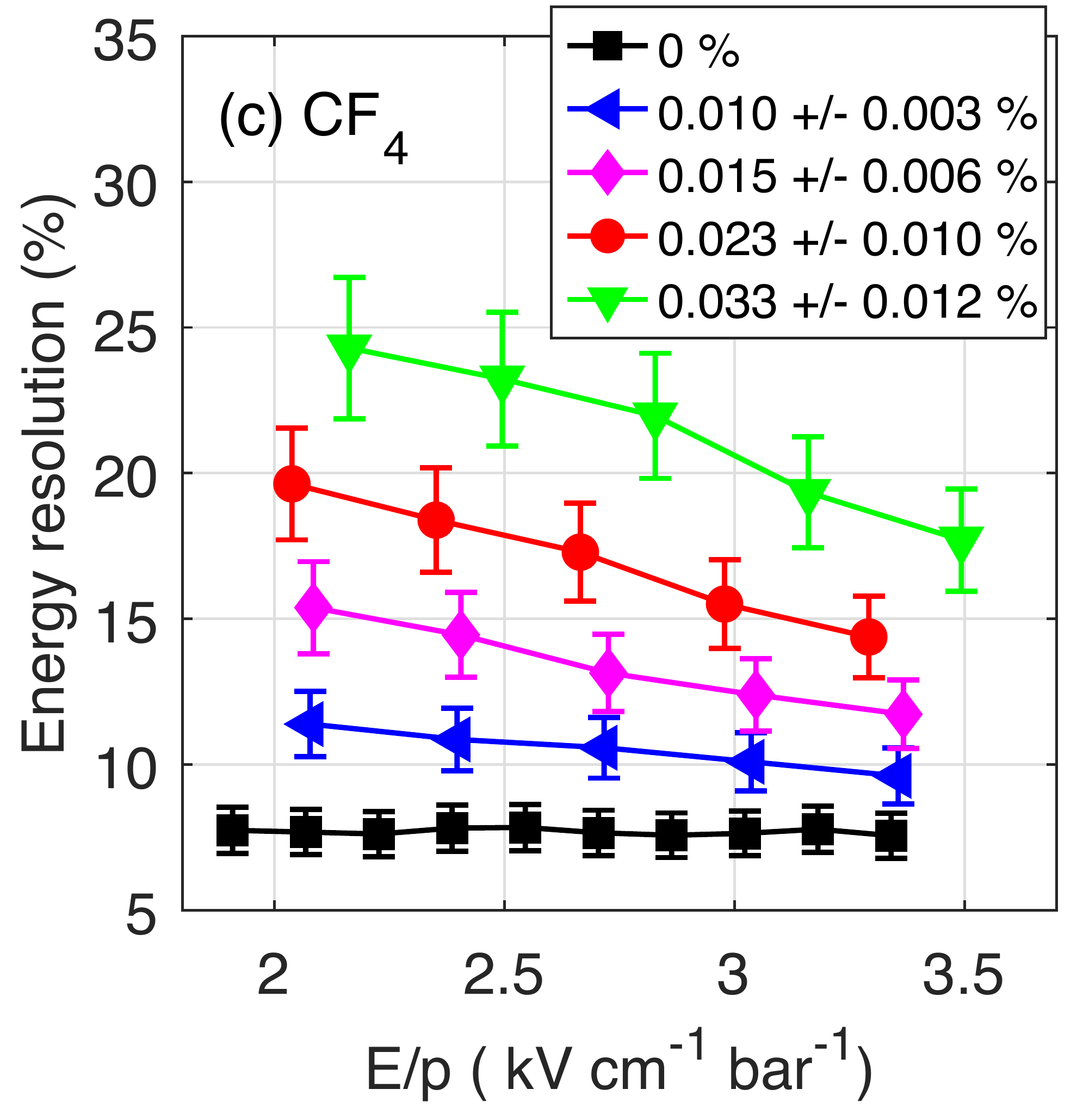}
\caption{\label{fig:resolution} Energy resolution for 5.9 keV x-rays absorbed in the driftless GPSC as a function of $E/p$, for mixtures of Xe with various molecular additives at different concentrations: (a) CO${}_{2}$; (b) CH${}_{4}$; (c) CF${}_{4}$. Total pressures of 1.13, 1.25 and 1.24 bar were used, in average, for Xe-CO${}_{2}$, Xe-CH${}_{4}$ and Xe-CF${}_{4}$ mixtures, respectively. The solid lines serve only to guide the eye.} 
\end{figure}

The energy resolution is crucial in neutrinoless double beta decay searches since it is the only known asset that allows a discrimination between $\beta\beta0\nu$ events against the $\beta\beta2\nu$ background. The impact of molecular additives in the energy resolution obtained in the xenon GPSC was evaluated for the Xe-CO${}_{2}$, Xe-CH${}_{4}$ and Xe-CF${}_{4}$ mixtures used in this work. Figure~\ref{fig:resolution} shows the energy resolution (FWHM) for 5.9 keV x-rays absorbed in the driftless GPSC as a function of the reduced electric field, for different additive concentrations. The error bars have a statistical contribution related to the confidence interval of the fit parameters, being strongly affected by the unknown attachment probability, which is left as free parameter in the fit. An additional systematic error was added, estimated from the response function fitting method developed in this work. As shown, the energy resolution for CO${}_{2}$ and CH${}_{4}$ mixtures does not degrade significantly up to 0.04\% and 0.4\% concentrations, respectively, in particular at high electric fields, and the best energy resolution is achieved with Xe-CH${}_{4}$ mixtures. For CF${}_{4}$, the energy resolution obtained is strongly deteriorated even at concentrations as low as 0.02\%, something that we attribute to dissociative attachment.

The energy resolution at zero-depth (i.e. corresponding to a 2.5 cm EL gap) allows for evaluation of the different contributions to the energy resolution, in particular the fluctuations in the number of EL photons produced. For that we recall that the energy resolution, $R$, of a GPSC can be described by~\cite{r25}:

\begin{equation}
\label{eq:R}
R=2.355\sqrt{\frac{F}{N_e}+\frac{Q}{N_e}+\frac{1}{N^{\ }_{pe}}\left(1+\frac{{\sigma}_{_G}^2}{{G}^2}\right)\ }.
\end{equation}

\noindent The first term under the square root accounts for the relative fluctuations in the number of ionisation electrons released by the interacting radiation, described by the Fano factor, defined as $F={{\sigma }^2_e}/{N_e}$, where ${N}_{e}$ is the average number of primary electrons and ${\sigma}_e$ is the respective standard deviation. The second term accounts for relative fluctuations associated to the number of EL photons produced in the scintillation region per primary electron, ${N}_{el}$ being its average number and $Q={({\sigma}_{el}}/{N_{el})^2}$ the square of the corresponding relative standard deviation. The last term describes the relative fluctuations in the photosensor signal, associated to the number of  photoelectrons released from the PMT photocathode (${N}_{pe}$ being its average number), which follows a Poisson distribution, and the relative fluctuations in the gain of the electron avalanche produced in the PMT dynodes (with ${G}$ and ${\sigma}_{_G}$ being the average gain and the corresponding standard deviation, respectively). The average number of photoelectrons released from the PMT photocathode is given by ${N}_{pe}=c {N_e} {N}_{el}$, $c$ being the light collection efficiency, which depends on the transparency of the anode grid and the PMT window, on the PMT quantum efficiency and on the average solid angle subtended by the PMT photocathode relative to the primary electron path in the EL region.

For pure xenon, the contribution from statistical fluctuations associated to the EL production (Q) is negligible when compared to the other factors~\cite{r25}, allowing an experimental determination of the energy resolution contributions resulting from statistical fluctuations in the primary ionisation formation and in the photosensor. In this case, equation~\eqref{eq:R} can be approximated by:

\begin{equation}
\label{eq:R2}
R=2.355 \sqrt{\frac{F}{N_e}+\frac{1}{c{N_e}{N}_{el}}\left(1+\frac{{\sigma}^2_{_G}}{{G}^2}\right)}.
\end{equation}

\noindent By plotting $R^2$ as a function of $1/{N}_{el}$, a linear function can be fitted to the data points. From the vertical intercept and the slope of the line, the Fano factor and photosensor contributions can be estimated, respectively, similarly to what happens in standard GPSCs with drift region~\cite{r34,r35}. Using pure xenon data, we have obtained a Fano factor $F=0.21 \pm 0.04$~\cite{r29}. This result is in agreement with the values normally found in the literature, between 0.13 and 0.25~\cite{r35, r36, r37, r38}. Moreover, the photosensor contribution obtained from the linear fit is compatible with calculations based on the geometry and the PMT characteristics in our setup, confirming the robustness of the method used in this work.

We assume that the contribution to the energy resolution from statistical fluctuations in the primary ionisation formation  are constant for the additive concentrations studied in this work, since the Fano factor and the $w$-values of those mixtures are not expected to change significantly at these very small additive concentrations. This assumption makes it possible to use equation~\eqref{eq:R} to determine the fluctuations associated to EL production. The Fano factor, $F$, and the term containing the photo-sensor statistics $1/c\left(1+{\sigma}^2_{_G}/{G}^2\right)$ are simply taken from the pure xenon fit.

\begin{figure}[tbp]
\centering 
\includegraphics[width=7.4cm,height=\textheight,keepaspectratio]{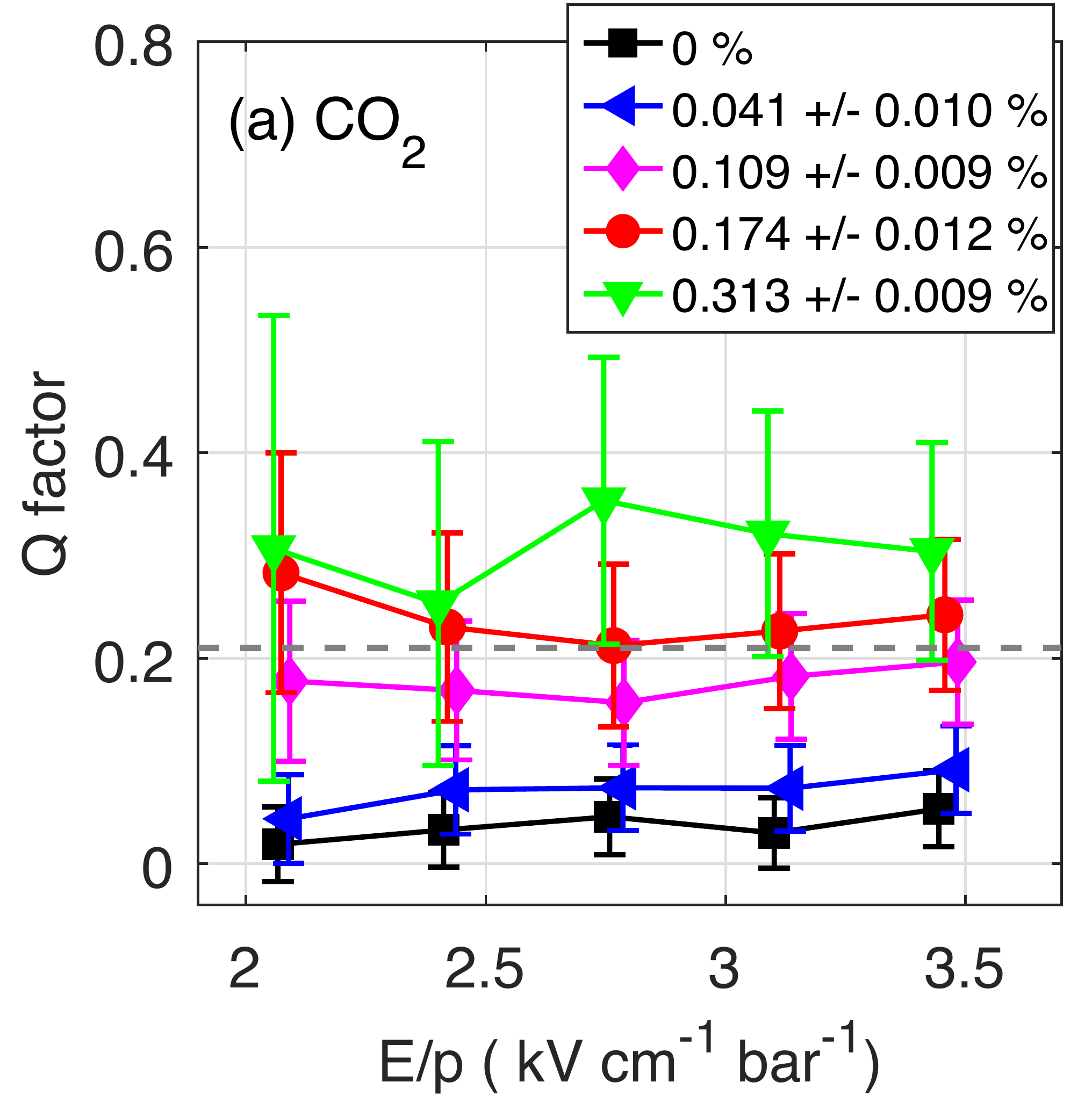}
\includegraphics[width=7.4cm,height=\textheight,keepaspectratio]{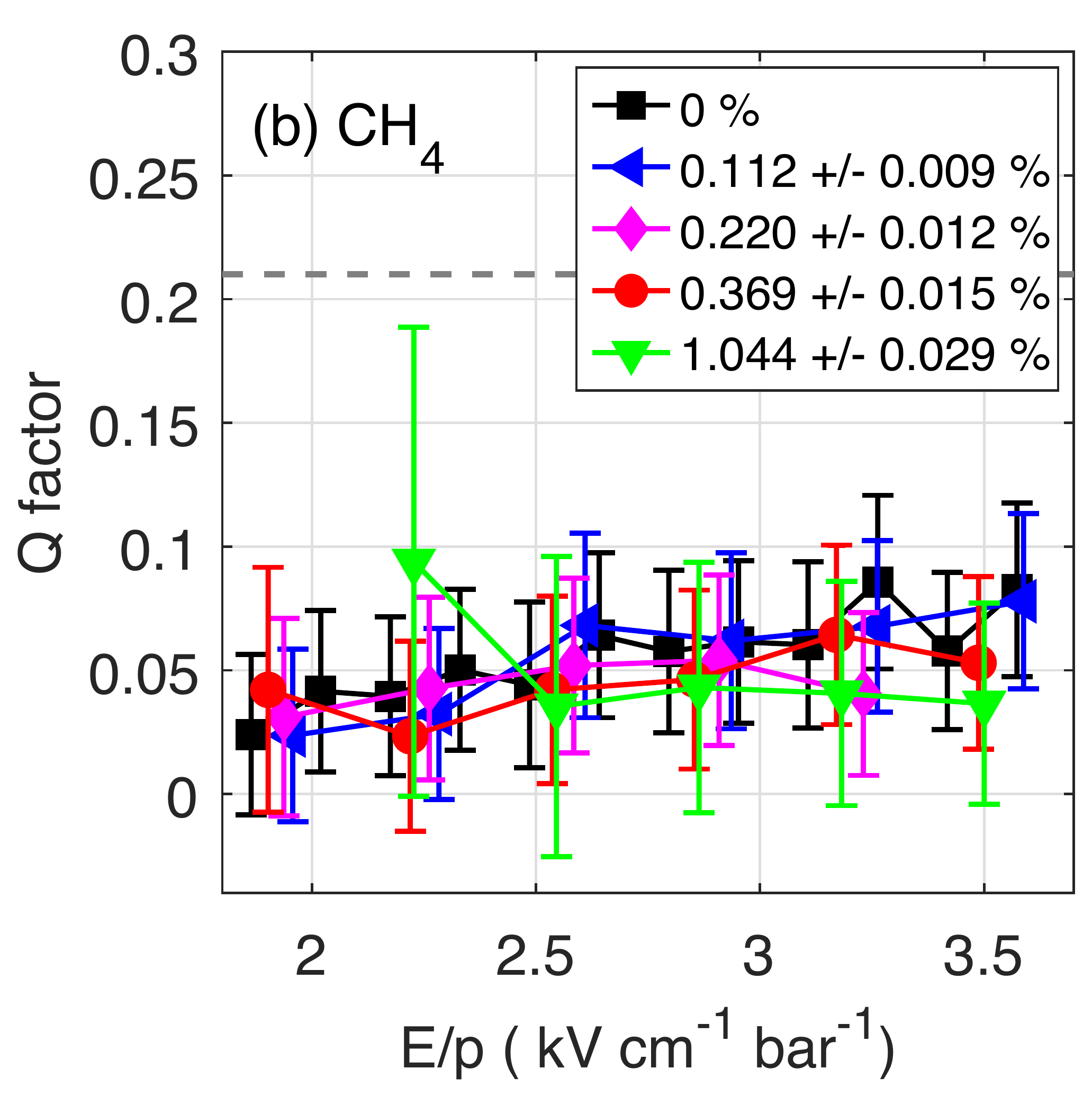}
\includegraphics[width=7.4cm,height=\textheight,keepaspectratio]{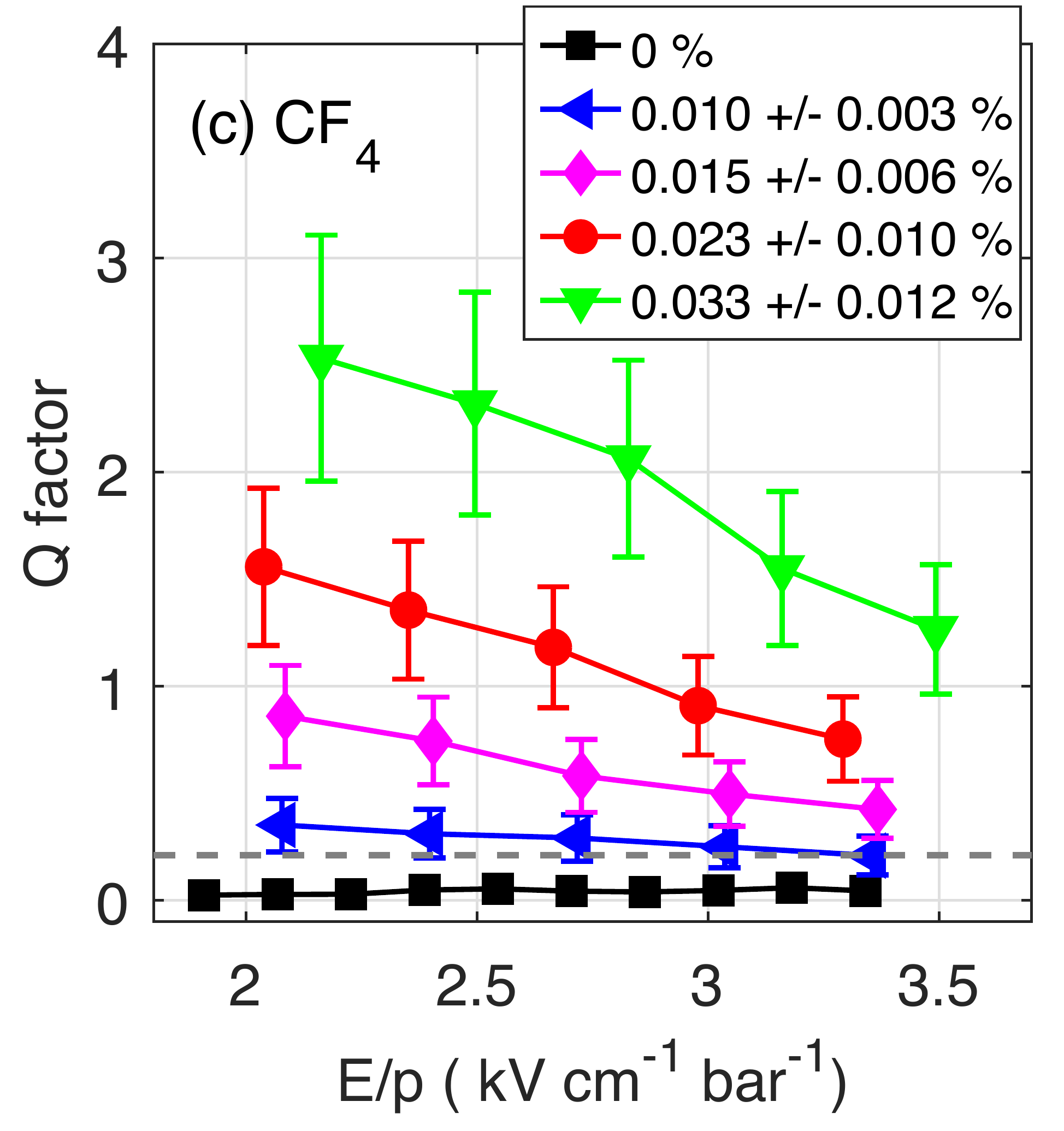} 
\caption{\label{fig:Q} Square of the relative standard deviation in the number of EL photons produced per primary electron ($Q$) as a function of $E/p$ in the scintillation region (25 mm wide) for different concentrations of molecular additives: (a) CO${}_{2}$; (b) CH${}_{4}$; (c) CF${}_{4}$. Total pressures of 1.13, 1.25 and 1.24 bar were used, in average, for Xe-CO${}_{2}$, Xe-CH${}_{4}$ and Xe-CF${}_{4}$ mixtures, respectively. The horizontal dashed line indicates the Fano factor. The solid lines serve only to guide the eye.}
\end{figure}

In figure~\ref{fig:Q} we present the square of the relative standard deviation in the number of EL photons produced in the scintillation region per primary electron, $Q={({\sigma}_{el}}/{N_{el})^2}$, as a function of $E/p$, in the range normally used in the scintillation region, for different concentrations of molecular additives: (a) CO${}_{2}$, (b) CH${}_{4}$, (c) CF${}_{4}$. The error bars come from the errors on the energy resolution values (figure~\ref{fig:resolution}) and from the two parameters of the linear fit to $R^2$ as a function of $1/{N}_{el}$, from which the Fano factor and the photosensor contributions were obtained (in pure xenon). As shown, $Q$ is negligible for pure xenon, when compared to $F$, and tends to increase as the additive concentration increases. For CO${}_{2}$, $Q$ is not strongly dependent on the reduced electric field and for concentrations up to 0.1\% it is still below the Fano factor. For CH${}_{4}$, $Q$ does not depend significantly on $E/p$ in the range considered. CH${}_{4}$ concentrations as high as 0.7\% result in negligible $Q$ values. This explains why the energy resolution obtained in Xe-CH${}_{4}$ mixtures is better when compared to Xe-CO${}_{2}$ and Xe-CF${}_{4}$ mixtures for the investigated range of additive concentrations. For CF${}_{4}$, a minute concentration of 0.01\% is enough to make $Q$ larger than $F$, rising abruptly as the concentration increases. Additive concentrations above the ones shown in figure~\ref{fig:Q} result in lower signal-to-noise ratios, which may worsen the energy resolution obtained, resulting in over-estimated $Q$ values from equation \eqref{eq:R} since the noise contribution is not included in that equation.

The rise in the $Q$ factor as the additive concentration increases cannot be explained if we take only into account the effect of EL reduction with increasing additive concentration. In addition, one would expect a decrease in $Q$ with increasing electric field in the scintillation region, which is not observed for CO${}_{2}$. As argued in~\cite{r32}, the effect can be interpreted as due to dissociative electron attachment. According to simulations, attachment is negligible for the Xe-CH${}_{4}$ mixtures investigated in this work, in particular at high electric fields. Its presence becomes nonetheless the main source of fluctuations in the EL signal for CO${}_{2}$ concentrations above 0.2\% and for CF${}_{4}$ concentrations above 0.01\%.

\section{Discussion}
\label{sec:discussion}

\begin{figure}[tbp]
\centering 
\includegraphics[width=\textwidth,height=4.9cm,keepaspectratio]{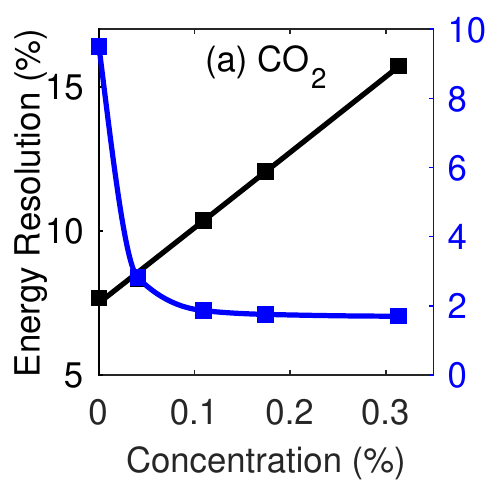}
\includegraphics[width=\textwidth,height=4.9cm,keepaspectratio]{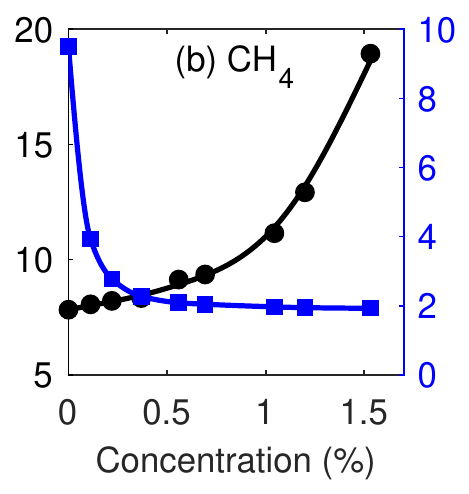}
\includegraphics[width=\textwidth,height=4.9cm,keepaspectratio]{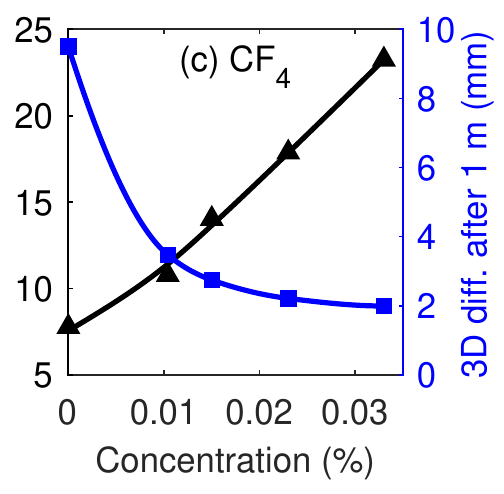} 
\caption{\label{fig:concentration} Energy resolution (in black) as a function of additive concentration, obtained under an electric field of 2.5 kV/cm/bar in the GPSC scintillation region, for xenon admixtures with: (a) CO${}_{2}$; (b) CH${}_{4}$; (c) CF${}_{4}$.  The corresponding 3D diffusion predicted from Magboltz simulations~\cite{r39} after 1 m drift is shown in the right hand vertical axis (in blue), and was obtained for a xenon TPC with a nominal pressure of 10 bar and a reduced electric field of 20 V/cm/bar in the drift region}
\end{figure}

The results described in previous section need to be put in the context of the effective reduction of electron diffusion in the NEXT-100 TPC for each mixture investigated. As starting point, we present in figure~\ref{fig:concentration} the energy resolution (in black) obtained for the three molecular additives investigated as a function of additive concentration, for typical operation conditions of the GPSC, in particular a reduced electric field of 2.5 kV/cm/bar in the scintillation region. As seen, the tendency of degradation of the energy resolution with the amount of additive is different for CH${}_{4}$ when compared to CO${}_{2}$ and CF${}_{4}$. In the first case, the energy resolution starts to degrade very slowly up to concentrations of 0.7\%, increasing faster above 1.0\%, while for CO${}_{2}$ and CF${}_{4}$ the variation is almost linear in the considered concentration range. In addition, we present in figure~\ref{fig:concentration} the 3D diffusion predicted after 1 m drift (in blue). Electron diffusion was simulated with Magboltz~\cite{r39} for the additive concentrations used in this work, for a xenon TPC with a nominal pressure of 10 bar and a reduced electric field of 20 V/cm/bar in the drift region. From the comparison between the obtained energy resolution and the expected 3D diffusion, the best compromise between low diffusion and minor energy resolution degradation favors the choice of CH${}_{4}$ as xenon additive and can be found at CH${}_{4}$ concentrations between 0.2\% an 0.5\%. 

The drastic change in the scintillation threshold observed in figure~\ref{fig:yield} suggests that electron cooling is strongly active for these minute concentrations of additives. Moreover, simulations indicate that a minimum of diffusion exists in the drift field range of 20-30 V/cm/bar~\cite{r32}. Although in pure xenon at 10 bar, drift fields in that range are not critical concerning recombination of the primary electrons, for xenon admixtures the situation at high pressures is not so clear, and should be carefully studied in the future. 

The EL yield and the energy resolution obtained for the three additives investigated were extrapolated to the expected operation conditions of the NEXT-100 TPC, including a nominal pressure of 10 bar and a reduced electric field of 2.5 kV/cm/bar in the EL region. Electron diffusion was obtained with Magboltz simulations~\cite{r39} for the additive concentrations used in this work and the operational conditions of the NEXT-100 TPC. In order to obtain the extrapolated energy resolution, we have used equation~\eqref{eq:R}, assuming a Fano factor $F=0.15$, a relative standard deviation in the PMT gain ${\sigma}_G/G=0.35$ and a light collection efficiency $c=0.01$. Sharp discontinuities of the Fano factor are expected to occur for x-ray energies near the Xe absorption edges\cite{r40}, resulting in a higher Fano factor, which is the case of the x-rays used in our study (5.9 keV is just above the Xe L-edges). Therefore, in the NEXT-100 extrapolation, we opted for a lower Fano factor of 0.15 since there are no measurements at 2.45 MeV and no sharp discontinuities are expected for electrons, keeping the consistency with previous studies performed by the NEXT collaboration\cite{r41,r42}. To extrapolate the number of EL photons we have considered a scintillation gap of 0.6 cm, a $w$-value of 22 eV and the energy of $\beta\beta0\nu$ events ${Q}_{\beta\beta}=$ 2.458 MeV. The ratio between the scintillation probabilities for 1 bar and 10 bar is taken from simulation, being largely dominated by the increased quenching probability of the Xe$_2^*$ triplet state at high pressure~\cite{r32}. We have assumed a 100\% transparency for CH${}_{4}$ and CF${}_{4}$, while for CO${}_{2}$ a correction for the amount of light lost to photo-absorption in 2 m (the maximum length intended for the NEXT-100 TPC and upgrades) was introduced. Concerning $Q$, for CO${}_{2}$ and CF${}_{4}$ the experimental values obtained in this work at about 1 bar were scaled considering the relation $Q \simeq \eta g/3$ derived earlier in~\cite{r32}, where $g$ is the EL gap and $\eta$ the dissociative attachment probability per unit of length, and further taking into account the effect of the solid angle variation along the electron path, which is present in our detector but not in the NEXT-100 TPC. For CH${}_{4}$, experimental $Q$ values obtained in this work were directly used since we do not expect that $Q$ changes significantly at 10 bar, according to simulation~\cite{r32}.

\begin{figure}[tbp]
\centering 
\includegraphics[width=11.5cm,height=\textheight,keepaspectratio]{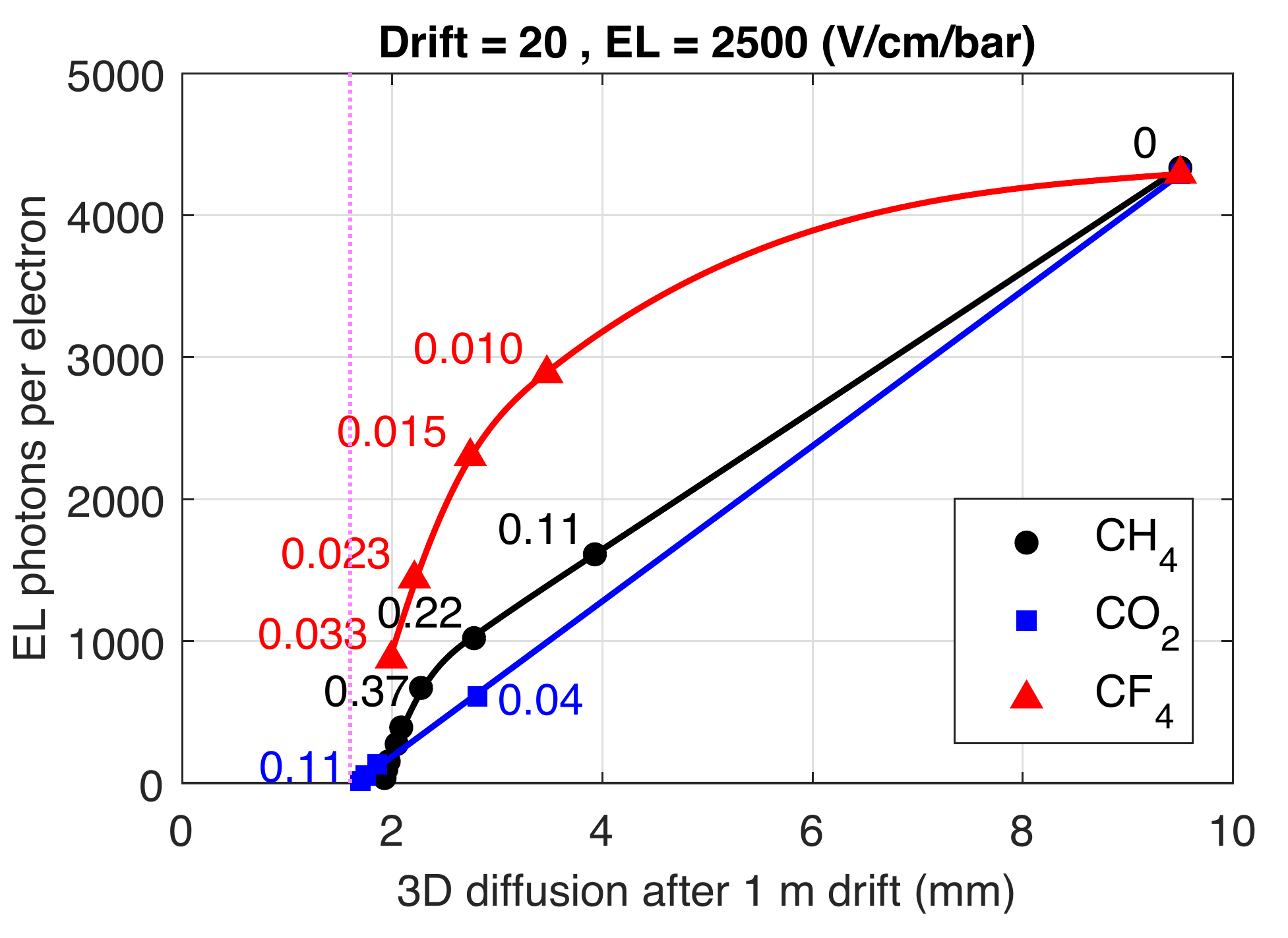}
\caption{\label{fig:YvD} Measured number of xenon EL photons, extrapolated to the NEXT-100 TPC operating conditions ($E/p=20$ V/cm/bar in the drift region and $E/p=2.5$ kV/cm/bar in the EL region, $p=10$ bar), as a function of the simulated 3D diffusion after 1 meter of drift for different concentrations (indicated in \%) of the molecular additives investigated in this work (CO${}_{2}$, CH${}_{4}$ and CF${}_{4}$). The vertical line indicates the thermal diffusion for the described conditions.}
\end{figure}

Figure ~\ref{fig:YvD} shows the number of xenon EL photons, extrapolated to typical operation conditions of the NEXT-100 TPC, as a function of the 3D diffusion coefficient, defined as ${D}_{3d}=\sqrt[3]{D_T\times D_T\times D_L}$, $D_T$ and $D_L$ being the transverse and longitudinal diffusion coefficients, respectively, obtained from Magboltz~\cite{r39} for different concentrations of the three additives investigated. The 3D diffusion coefficient is defined here as the characteristic size of the electron diffusion ellipsoid ($\sqrt[3]{xyz}$) after 1 m drift through the TPC. For CO${}_{2}$, the number of EL photons was corrected for the transparency to VUV photons in the NEXT-100 TPC, dropping from about 65\% to 23\% as the CO${}_{2}$ concentration increases from 0.04\% to 0.31\%. As diffusion decreases, the drop in the EL yield is much smaller for CF${}_{4}$ when compared to CO${}_{2}$ and CH${}_{4}$. Still, the drop in EL yield for CO${}_{2}$ and CH${}_{4}$ seems tolerable, provided the number of photons produced per ionisation electron is large enough and the electric field intensity in the scintillation region can be increased. To reach an electron diffusion in the range 2.5-3.0 mm/$\sqrt{\mathrm{m}}$, one needs additive concentrations of about 0.04\%, 0.2\% and 0.015\% for CO${}_{2}$, CH${}_{4}$ and CF${}_{4}$, respectively. 

\begin{figure}[tbp]
\centering 
\includegraphics[width=11.5cm,height=\textheight,keepaspectratio]{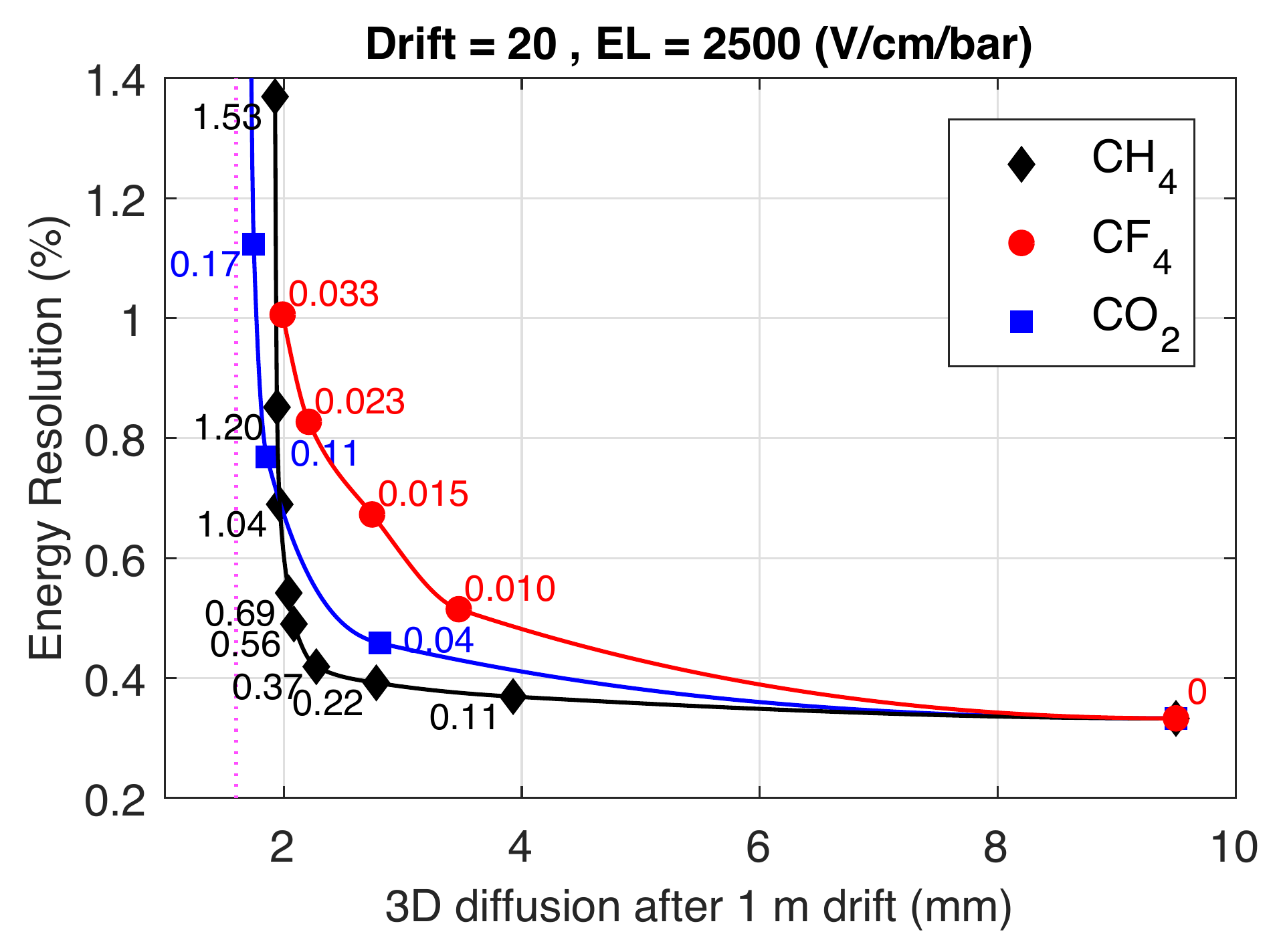}
\caption{\label{fig:RvD} Energy resolution extrapolated to the energy of a $\beta\beta0\nu$ decay event (2.5 MeV) and to the NEXT-100 TPC operating conditions ($E/p=20$ V/cm/bar in the drift region and $E/p=2.5$ kV/cm/bar in the EL region), as a function of the simulated 3D diffusion after 1 meter of drift, for different concentrations (indicated in \%) of the molecular additives investigated in this work (CO${}_{2}$, CH${}_{4}$ and CF${}_{4}$). The vertical line indicates the thermal diffusion for the described conditions.}
\end{figure}

Figure~\ref{fig:RvD} shows the energy resolution measured for 5.9 keV x-rays, extrapolated to the energy of a $\beta\beta0\nu$ decay event (2.5 MeV) and typical operating conditions in the NEXT-100 TPC ($c=0.01$ and ${\sigma}_G/G=0.35$), as a function of the 3D diffusion (after 1 m drift), obtained from Magboltz~\cite{r39} for different concentrations of the three additives investigated, and for reduced electric fields of 20 V/cm/bar in the drift region and 2.5 kV/cm/bar in the scintillation region. As seen, the best compromise between energy resolution and diffusion is found for concentrations of about 0.04\%, 0.4\% and 0.01\% for CO${}_{2}$, CH${}_{4}$ and CF${}_{4}$, respectively. The best energy resolution obtained in these conditions favours CH${}_{4}$ as a choice. Despite the high quenching for this additive, an energy resolution of 0.5\% at ${Q}_{\beta\beta}$ can be obtained for ${D}_{3d}=2$ mm (CH${}_{4}$ concentration of 0.5\%). For CO${}_{2}$, the energy resolution is also not much deteriorated but its performance is affected by transparency and getter compatibility. For CF${}_{4}$, quenching is low but the energy resolution is worse due to the high electron attachment. 

There are some aspects that need to be evaluated at high pressures in a larger prototype, in particular the pressure-dependence of the primary and secondary scintillation yields and the fluctuations in light production, together with experimental studies on the electron diffusion that can be reached in those conditions, and the impact of charge recombination (if any). The NEXT-DEMO prototype, with a drift length of 30 cm, operated with 1.5 kg of natural xenon at a pressure of 10 bar, has the potential to study all the above mentioned aspects~\cite{r43}. Despite the anticipated performance deterioration at high pressure, the extrapolations performed here, based on the simulation package introduced in~\cite{r32}, indicate that both the $Q$ factor and fluctuations in the PMT signal can be kept at the level of the Fano factor at 10 bar, for additive concentrations in the optimum range for ${D}_{3d}$ (2 to 3 mm/$\sqrt{\mathrm{m}}$). Concerning the primary scintillation yield, a tolerable reduction of about 80\% can be extrapolated for CH${}_{4}$ and CO${}_{2}$ in the same concentration region. Due to difficulties inherent to instability of long-term operation of Xe-CO${}_{2}$ mixtures in the presence of getters, together with the good performance demonstrated by Xe-CH${}_{4}$ mixtures, CH${}_{4}$ seems to be the best molecular additive to use in the NEXT-100 xenon TPC.

\section{Conclusions}
\label{sec:conclusion}

We have demonstrated that the addition of molecular gases to pure xenon, at sub-percent concentration levels, is not dramatic in terms of electroluminescence yield, as it has been assumed over the last decades. Comparing the EL yield obtained as a function of the expected electron diffusion for the three additives investigated, we observe that the 3D diffusion coefficient diminishes from 10 to about 2 mm/$\sqrt{\mathrm{m}}$ with the drop in EL being much less significant for CF${}_{4}$ than for CO${}_{2}$ and CH${}_{4}$. 

The intrinsic energy resolution of xenon-based TPCs degrades with increasing additive concentration as well. Up to additive concentrations of about 0.04\% for CO${}_{2}$ and 0.4\% for CH${}_{4}$, there is no significant degradation at $E/p$ values above 2.5 kV/cm/bar. For those concentrations, the contribution of the statistical fluctuations associated to EL production ($Q$ factor) is lower than the Fano factor. Fluctuations are however higher for CO${}_{2}$ when compared to CH${}_{4}$ due to dissociative electron attachment by CO${}_{2}$ molecules. The energy resolution obtained for CF${}_{4}$ is much worse than that obtained for the two other molecular additives due to a much higher level of this latter process.

Comparing the results obtained for the three candidate additives investigated, after extrapolating to the operation conditions of the NEXT-100 TPC, it is clear that CF${}_{4}$ cannot be used as a result of the huge fluctuations observed in the EL formation, yielding an extremely high $Q$ factor. Furthermore, the concentration levels of CF${}_{4}$ would be very small and difficult to handle for several reasons. The measurement of the CF${}_{4}$ concentration itself is difficult since we are close to the RGA sensitivity. Another technical difficulty is the preparation of the mixture as part of the CF${}_{4}$ molecules are adsorbed (to the walls for example), which would affect the final concentration.

The comparison between CO${}_{2}$ and CH${}_{4}$ favours the latter. CH${}_{4}$ does not present the drawback of having significant electron attachment, but on the other hand displays a higher quenching. As a result, the EL yield is comparable for both cases, but the energy resolution is considerably better for CH${}_{4}$. Higher concentration levels of CH${}_{4}$ are however needed in order to efficiently reduce electron diffusion, which would have a high impact on the discrimination of events through pattern recognition of the topology of primary ionisation trails. CO${}_{2}$ has additional disadvantages of not being 100\% transparent to VUV photons in a large chamber (NEXT-100), and presenting long-term instability in the presence of getters. CH${}_{4}$ has proven to be the best candidate for the NEXT-100 TPC but additional studies are needed in a larger TPC (NEXT-DEMO). These studies include chiefly the evaluation of the pressure-dependence of the primary and secondary scintillation yields and the EL fluctuations, together with the determination of the electron diffusion that can be achieved without significantly compromising the TPC calorimetric performance.

\acknowledgments
The NEXT Collaboration acknowledges support from the following agencies and institutions: the European Research Council (ERC) under the Advanced Grant 339787-NEXT; the European Union’s Framework Programme for Research and Innovation Horizon 2020 (2014-2020) under the Marie Skłodowska-Curie Grant Agreements No. 674896, 690575 and 740055; the Ministerio de Econom\'ia y Competitividad of Spain under grants FIS2014-53371-C04, the Severo Ochoa Program SEV-2014-0398 and the Mar\'ia de Maetzu Program MDM-2016-0692; the GVA of Spain under grants PROMETEO/2016/120 and SEJI/2017/011; the Portuguese FCT under project PTDC/FIS-NUC/2525/2014, under project UID/FIS/04559/2013 to fund the activities of LIBPhys, and under grants PD/BD/105921/2014, SFRH/BPD/109180/2015 and SFRH/BPD/76842/2011; the U.S.\ Department of Energy under contracts number DE-AC02-07CH11359 (Fermi National Accelerator Laboratory), DE-FG02-13ER42020 (Texas A\&M) and DE-SC0017721 (University of Texas at Arlington); and the University of Texas at Arlington. DGD acknowledges Ramon y Cajal program (Spain) under contract number RYC-2015-18820. We also warmly acknowledge the Laboratori Nazionali del Gran Sasso (LNGS) and the Dark Side collaboration for their help with TPB coating of various parts of the NEXT-White TPC. Finally, we are grateful to the Laboratorio Subterr\'aneo de Canfranc for hosting and supporting the NEXT experiment.


\end{document}